\documentclass[letterpaper]{article}
\usepackage{times}
\usepackage{amsmath,amssymb,graphicx}
\usepackage{latexsym}

\begin{document}
\title{A New Generalized Closed Form Expression for Average Bit Error Probability Over Rayleigh Fading Channel}
\author{Sanjay Singh \thanks{Sanjay Singh is with the Department of Information and Communication Technology, Manipal Institute of Technology, Manipal University, Manipal-576104, INDIA, E-mail: sanjay.singh@manipal.edu},~M.~Sathish Kumar and Mruthyunjaya H.S \thanks{M.~Sathish Kumar and Mruthyunjaya H.S are with, 
Department of Electronics and Communication Engineering, Manipal Institute of Technology, Manipal-576104, INDIA}}
\maketitle

\begin{abstract}
Except for a few simple digital modulation techniques, derivation of average bit error probability over fading channels is difficult and is an involved process. In this letter, curve fitting technique has been employed to express bit error probability over AWGN of any digital modulation scheme in terms of a simple Gaussian function. Using this Gaussian function, a generalized closed form expression for computing average probability of bit error over Rayleigh fading channels has been derived. Excellent agreement has been found between error probabilities computed with our method and the rigorously calculated error probabilities of several digital modulation schemes.
\end{abstract}

\section{Introduction}
Bit Error Rate (BER) or bit error probability is one of the critical performance measures of digital communication systems. In elementary systems the channel can be modeled as an AWGN wherein the calculation of BER is straightforward. However, for mobile environments, since the signal is received via multipath with associated fading, the BER of AWGN channels is no longer valid. 
\par
To compute the average BER of a given modulation scheme in fading channels, the corresponding BER of that modulation scheme over AWGN is averaged over the fading statistics of the channel. While the procedure is seemingly simple, one often encounters major difficulties while averaging the AWGN BER over fading channels. 
\par
To simplify the derivation of average bit error probability over fading channels, there are so many approaches in the literature for approximating Q-function  \cite{ta2000} \cite{cd02} \cite{si02} \cite{kl07} \cite{rm08} \cite{dd08} \cite{ir08} and more recently \cite{cb09}. In this paper, instead of direct approximation of Gaussian Q-function, we approximate to a high degree of accuracy, the BER \textit{itself} of any arbitrary digital modulation technique by a Gaussian function. This leads us to an exact analytical closed form expression for the average bit error probability of any given digital modulation technique.
\section{System model}
Assuming absence of line of sight, the received signal r(t) in a Rayleigh fading channel is expressed as
\begin{equation}\label{eq:1}
r(t)=h(t)s(t)+n(t)
\end{equation}

where $h$: is the complex scaling factor corresponding to the Rayleigh multipath channel\\
      $s$: is the transmitted signal\\
      $n$: is the sample function of White Gaussian Noise with mean $\mu$ and variance $No/2$.
  Following assumptions have been made\\
  \begin{enumerate}
  \item The channel is flat
  \item The channel is randomly varying in time with each transmitted symbol getting multiplied by a randomly varying complex quantity $h$. Since $h$ is modeling a Rayleigh channel, the real and imaginary parts are Gaussian distributed with zero mean and variance 1/2.
   \end{enumerate}
 \section{Average bit error probability computation over the Rayleigh fading channel}
Let $P_ {b,GDMT}^{AWGN}$ stand for the probability of bit error for a General Digital Modulation Technique (GDMT) over an AWGN channel. The GDMT can be any digital modulation technique such as PSK, QPSK, QAM, FSK, ASK etc. Let $P_{b,GDMT}^{Fading}$ denote the average probability of bit error for the GDMT over a Rayleigh fading channel.  In the following section we first discuss the Gaussian function modeling of $P_{b,GDMT}^{AWGN}$ which will be followed by derivation of an expression for $P_{b,GDMT}^{Fading}$ based on this Gaussian function model . 

\subsection{Approximation of bit error probability $P_{b,GDMT}^{AWGN}$ in terms of Gaussian function}
We model the BER probability of a GDMT over an AWGN as a Gaussian function using simple curve fitting techniques. Towards this, we first express $P_ {b,GDMT}^{AWGN}$ as given below:
\begin{equation}\label{eq:2}
 P_{b,GDMT}^{AWGN}=a\exp\big[-\big(\frac{x-b}{c}\big)^2\big] 	
 	\end{equation}
Where a, b and c are constants to be determined through simple curve fitting techniques. For the curve fitting, the following parameters have been used to measure the goodness of fit \cite{br03}:
	\begin{itemize}
	\item Sum of Squares due to Errors (SSE) : A value closer to 0 indicates a better fit.
	\item R-square : A value closer to 1 indicates a better fit.
	\item Adjusted R-square : Any value less than or equal to 1 indicates a better fit.
	\item Root-Mean Square Error (RMSE) : A value closer to 0 indicates better fit.
	\end{itemize}
Depending on the range taken by $x$ there can be an infinite number of choices for the values of a, b, and c. Since values of practical interest in $E_b/N_o$ range from 0 to only a few tens of dB, one needs to obtain the values of a, b, and c, only in this range of $E_b/N_o$ and not the entire $x$ axis. The values of constants $a$, $b$ and $c$ along with the values of goodness of fit for several digital modulation techniques is given in Table~\ref{tab:1} and \ref{tab:2}.
\begin{table}[bpht]
\centering
\caption{Parametric Values for Approximating Function}
\label{tab:1}
\begin{tabular}{|c|c|c|c|}
\hline
Modulation Technique&$a$&$b$&$c$ \\ \hline
QPSK&0.1059&-2.405&4.344 \\ \hline
16-QAM&0.1793&0.3892&8.667 \\ \hline
BFSK&0.2036&-3.056&6.159 \\ \hline
BASK&0.1059&-2.405&4.344\\ \hline
\end{tabular}
\end{table}

\begin{table}[bpht]
\centering
\caption{Curve Fitting Performance Parameter Values}
\label{tab:2}
\begin{tabular}{|c|c|c|c|c|}
\hline
Modulation Technique&SSE&R-Square&Adjusted R-Square&RMSE\\ \hline
QPSK&$2.093\times 10^{-6}$&0.9998&0.9998&0.0002734 \\ \hline
16-QAM&$3.416\times 10^{-4}$&0.9978&0.9978&0.002668\\ \hline
BFSK&$2.169\times 10^{-5}$&0.9996&0.9996&0.0006722 \\ \hline
BASK&$2.093\times 10^{-6}$&0.9998&0.9998&0.0002088 \\ \hline
\end{tabular}
\end{table}

It has to be specifically noted that this simple form of Gaussian function modeling avoids the averaging of Q-function or error function squared over the channel fading statistics for certain types of digital modulation schemes. 
\subsection{BER Computation for GDMT over the Rayleigh fading channel}
In (2), the value of $x$ corresponds to $E_b/N_0$  or $|h|^2E_b/N_0$ for a pure AWGN or a fading channel respectively.  Now the conditional bit error probability of a GDMT for a sample value of the random variable $h$ is expressed as
\begin{equation}\label{eq:3}
P_{b|h}=a\exp\big[-\big(\frac{\xi-b}{c}\big)^2\big]
\end{equation}
with $\xi=|h|^2E_b/N_0$.  The average bit error probability of a GDMT over a fading channel is now computed by averaging the conditional probability given in (3) over the statistical distribution of $|h|^2E_b/N_0$ . Note that since $h$ has a Rayleigh distribution, $\xi$ will have a Chi squared distribution with degree of freedom two with pdf as given below \cite{pp08}   
\begin{equation}\label{eq:4}
	p(\xi)=\frac{1}{\gamma}\exp(-\frac{\xi}{\gamma}) \qquad \xi>0
	\end{equation}
where $\gamma=E[|h|^2]E_b/N_0$ with $E[.]$ being the expectation operator.
The average bit error probability, $P_{b,GDMT}^{Fading}$  over the fading channel can now be computed \cite{pr01} as 
	\begin{equation}\label{eq:5}
	P_{b,GDMT}^{Fading}=\int_{0}^{\infty}P_{b|h}(\xi)p(\xi)d\xi
	\end{equation}

On substituting $P_{b|h}(\xi)$ and $p(\xi)$ from (3) and (4) respectively in (5), we get
	
		\begin{equation}\label{eq:6}
		P_{b,GDMT}^{Fading}=\int_{0}^{\infty}ae^{-\big(\frac{\xi-b}{c}\big)^2}\frac{1}{\gamma}e^{-\frac{\xi}{\gamma}}d\xi
		\end{equation}
On solving (6), we get
\small
\begin{equation}\label{eq:7}
P_{b,GDMT}^{Fading}=\frac{ac\sqrt{\pi}}{2\gamma}\exp\bigg[\frac{-b^2}{c^2}+\frac{(\frac{2b}{c^2}-\frac{1}{\gamma})^2c^2}{4}\bigg]\bigg[1+\textit{\mbox{erf}}(\frac{b}{c}-\frac{c}{2\gamma})\bigg]
\end{equation}
where $\textit{\mbox{erf}}(.)$ is the error function and is defined as \cite{xi06} 
\begin{equation}\label{eq:8}
\textit{\mbox{erf}}(x)=\frac{2}{\sqrt{\pi}}\int_{0}^{x}\exp(-u^2)du.
\end{equation}

Hence, (7) gives a generalized closed form expression for the computation of average bit error probability over the Rayleigh fading channel for any digital modulation technique. It is worth mentioning at this stage that though in principle, the average error probability under fading can be computed, the process can be lengthy, and needs to be repeated when the modulation scheme changes, and  the equations need to be re-derived.
However, the novelty and utility of our work lies in the fact that using a simple Gaussian approximation of AWGN BER, the average error probability under Rayleigh fading for any given digital modulation scheme can be computed through a single expression, namely equation (7). 
\section{Exact average bit error probability computation over the Rayleigh fading channel}
To verify the validity of our generalized closed form expression (7) for BER over fading channels, we consider four common and practically used digital modulation techniques namely QPSK, M-QAM, M-FSK and M-ASK. The BER for these modulation techniques over AWGN can be found in the literature, for example in \cite{xi06}. Following the approach of section III(B) and using Craig's \cite{cr91} expression for the Q function as given below,

	\begin{equation}\label{eq:9}
		Q(x)=\frac{1}{\pi}\int_{0}^{\pi/2}\exp(-\frac{x^2}{2\sin^2\theta})d\theta
		\end{equation}
the rigorous expressions for the average BER of the various digital modulation techniques considered above can be arrived at and is as given below.

	\begin{equation}\label{eq:10}
	P_{b,QPSK}^{Fading}=\frac{1}{2}\Bigg(1-\sqrt{\frac{\gamma}{\gamma+1}}\Bigg)
	\end{equation}
	
	\begin{equation}\label{eq:11}
	P_{b,M-QAM}^{Fading}=K_1+K_2\sqrt{\frac{\beta_1\gamma}{\beta_1\gamma+2}}
	\end{equation}
	
	where $K_1=\frac{2\alpha_1-\alpha_1^2}{\log_2M}$, $K_2=\frac{4\alpha_1^2\arctan\Big(\sqrt{\frac{\beta_1\gamma+2}{\beta_1\gamma}}\Big)-2\pi\alpha_1}{\pi\log_2M}$ \\
	$\alpha_1=\frac{\sqrt{M}-1}{\sqrt{M}}$ and $\beta_1=\frac{3}{M-1}$
	
	\begin{equation}\label{eq:12}
	P_{b,M-FSK}^{Fading}=\frac{M}{2}\Bigg(1-\sqrt{\frac{\gamma\log_2M}{\gamma\log_2M+2}}\Bigg)
	\end{equation}
	and
	
	\begin{equation}\label{eq:13}
	P_{b,M-ASK}^{Fading}=\frac{\alpha_2}{2}\Bigg(1-\sqrt{\frac{\beta_2\gamma}{\beta_2\gamma+2}}\Bigg)
	\end{equation}
	where $\alpha_2=\frac{2(M-1)}{M\log_2M}$ and $\beta_2=\frac{6\log_2M}{M^2-1}$.
	\begin{figure}[bpht!]
\centering
\includegraphics[scale=0.7]{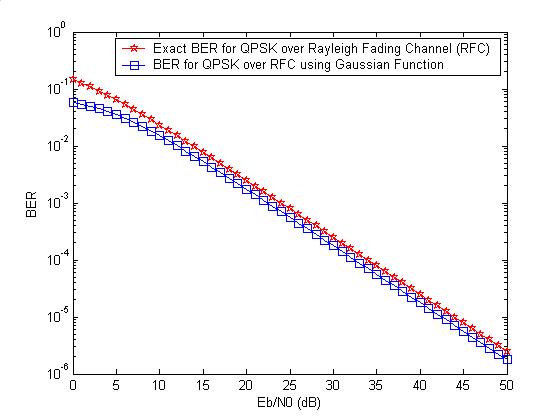}
\caption{Comparison of generalized and exact result for average BER computation over the Rayleigh fading channel for QPSK}
\label{fig:1}
\end{figure}

\begin{figure}[bpht!]
\centering
\includegraphics[scale=0.7]{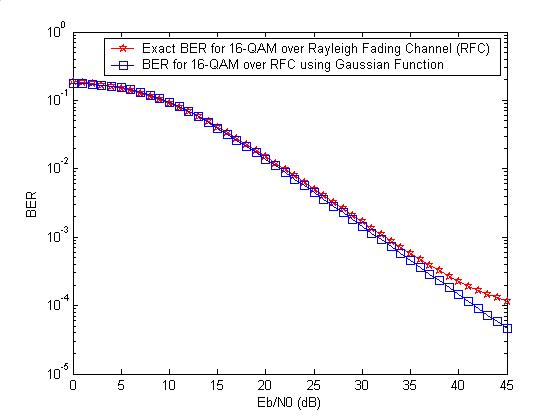}
\caption{Comparison of generalized and exact result for average BER computation over the Rayleigh fading channel for 16-QAM}
\label{fig:2}
\end{figure}

\begin{figure}[bpht!]
\centering
\includegraphics[scale=0.7]{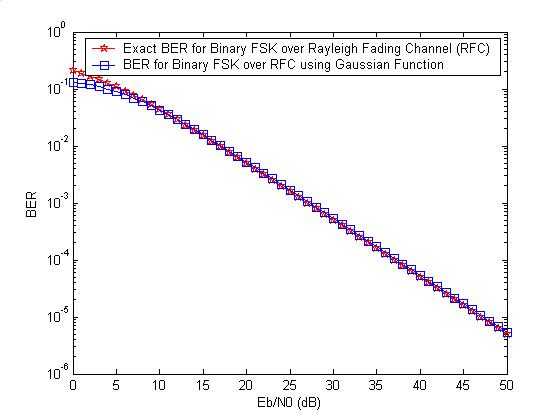}
\caption{Comparison of generalized and exact result for average BER computation over the Rayleigh fading channel for BFSK}
\label{fig:3}
\end{figure}

\begin{figure}[bpht!]
\centering
\includegraphics[scale=0.7]{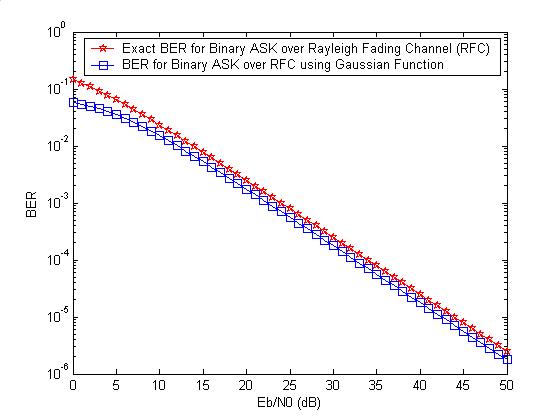}
\caption{Comparison of generalized and exact result for average BER computation over the Rayleigh fading channel for BASK}
\label{fig:4}
\end{figure}

\section{Verification of generalized result with the exact analytical results for various modulation techniques}
In the numerical results to be presented in this section, we have considered $E_b/N_0$ to range from 0 to 50 dB. The comparison of average BER results obtained through our generalized closed form expression and through the exact analytical expressions for the four digital modulation techniques considered is as shown in Figs \ref{fig:1} to \ref{fig:4}. Figure \ref{fig:1} and \ref{fig:2} shows the results for QPSK and 16-QAM while Fig \ref{fig:3} and \ref{fig:4} shows the results for FSK and ASK. These results clearly show the excellent agreement between the error probabilities computed with our method and the rigorously derived exact error probabilities.

\section{Conclusion}
A closed form expression for computing the bit error probabilities of any digital modulation technique in a Rayleigh fading channel has been derived. Our expression for the average error probability does not involve any numerical methods except simple curve fitting techniques and is based on representing the AWGN BER in the form of a general Gaussian function.  Numerical results show excellent agreement between error probability computed through our closed form expression and those obtained through rigorous derivation.
\bibliographystyle{IEEEtran}
\bibliography{ref}

\begin{thebibliography}{10}
\providecommand{\url}[1]{#1}
\csname url@samestyle\endcsname
\providecommand{\newblock}{\relax}
\providecommand{\bibinfo}[2]{#2}
\providecommand{\BIBentrySTDinterwordspacing}{\spaceskip=0pt\relax}
\providecommand{\BIBentryALTinterwordstretchfactor}{4}
\providecommand{\BIBentryALTinterwordspacing}{\spaceskip=\fontdimen2\font plus
\BIBentryALTinterwordstretchfactor\fontdimen3\font minus
  \fontdimen4\font\relax}
\providecommand{\BIBforeignlanguage}[2]{{%
\expandafter\ifx\csname l@#1\endcsname\relax
\typeout{** WARNING: IEEEtran.bst: No hyphenation pattern has been}%
\typeout{** loaded for the language `#1'. Using the pattern for}%
\typeout{** the default language instead.}%
\else
\language=\csname l@#1\endcsname
\fi
#2}}
\providecommand{\BIBdecl}{\relax}
\BIBdecl

\bibitem{ta2000}
C.Tellambura and A.Annamalai, ``Efficient computation of erfc(x) for large
  arguments,'' \emph{IEEE Trans.Commun.}, vol.~48, no.~4, pp. 529--532, April
  2000.

\bibitem{cd02}
M.~Chiani and D.~Dardari, ``Improved exponential bounds and approximation for
  the {Q}-function with application to average error probability computation,''
  in \emph{Global Telecommunications Conference, 2002. GLOBECOM '02. IEEE},
  vol.~2, nov. 2002, pp. 1399 -- 1402.

\bibitem{si02}
M.~K.Simon, ``Single integral representations of certain integer powers of the
  {G}aussian {Q}-function and their application,'' \emph{IEEE Commun.Lett.},
  vol.~6, no.~12, pp. 532--534, December 2002.

\bibitem{kl07}
G.~K.Karagiannidis and A.~S.Lioumpas, ``An improved approximation for the
  {G}aussian {Q}-function,'' \emph{IEEE Commun.Lett.}, vol.~11, no.~8, pp.
  644--646, August 2007.

\bibitem{rm08}
R.~M.Radaydeh and M.~M.Matalgah, ``Results for infinite integrals involving
  higher-order powers of the {G}aussian {Q}-function with application to
  average,'' \emph{IEEE Tran.Wireless Commun.}, vol.~7, no.~3, pp. 793--798,
  March 2008.

\bibitem{dd08}
J.~S.Dyer and S.~A.Dyer, ``Corrections to, and comments on, "an improved
  approximation for the {G}aussian {Q}-function,'' \emph{IEEE Commun.Lett.},
  vol.~12, no.~4, p. 231, April 2008.

\bibitem{ir08}
Y.Isukapalli and B.~D.Rao, ``An analytically tractable approximation for the
  {G}aussian {Q}-function,'' \emph{IEEE Commun.Lett.}, vol.~12, no.~9, pp.
  669--671, September 2008.

\bibitem{cb09}
Y.~Chen and N.~C.Beaulieu, ``A simple polynomial approximation to the
  {G}aussian {Q}-function and its application,'' \emph{IEEE Commun.Lett.},
  vol.~13, no.~2, pp. 124--126, February 2009.

\bibitem{br03}
P.~R.Bevington and D.~Robinson, \emph{Data Reduction and Error Analysis for the
  Physical Sciences}.\hskip 1em plus 0.5em minus 0.4em\relax New York:
  McGraw-Hill, 2003.

\bibitem{pp08}
A.Papoulis and S.~Pillai, \emph{Probability, Random Variables and Stochastic
  Processes}, 4th~ed.\hskip 1em plus 0.5em minus 0.4em\relax New Delhi: Tata
  McGraw-Hill, 2008.

\bibitem{pr01}
J.~G.Proakis, \emph{Digital Communications}, 4th~ed.\hskip 1em plus 0.5em minus
  0.4em\relax New Delhi: McGraw-Hill, 2001.

\bibitem{xi06}
F.~Xiong, \emph{Digital Modulation Techniques}, 2nd~ed.\hskip 1em plus 0.5em
  minus 0.4em\relax Boston: Artech House, 2006.

\bibitem{cr91}
J.~Craig, ``A new, simple and exact result for calculating the probability of
  error for two-dimensional signal constellations,'' in \emph{Military
  Communications Conference, 1991. MILCOM '91, Conference Record, 'Military
  Communications in a Changing World'., IEEE}, vol.~2, nov 1991, pp. 571 --575.

\end{thebibliography}

\end{document}